\documentclass[aps,prl,amsmath,amssymb,preprintnumbers,superscriptaddress,twocolumn,showpacs,floatfix]{revtex4}
\usepackage{epsfig,graphics,color}
\usepackage{graphicx}
\usepackage{dcolumn}
\usepackage{bm}
\usepackage{amsmath}

\newcommand \tie {{\it i.e.}}

\newcommand \kd  {\delta}
\newcommand \ra  {\rightarrow}

\newcommand \vecr {\vec{r}}

\newcommand \g {\gamma}

\newcommand \x {\cdot}

\newcommand \A {\alpha}

\newcommand \lc {\langle}
\newcommand \rc {\rangle}

\newcommand \D {\Delta}
\newcommand \sg {\sigma}

\newcommand \nt {\noindent}

\newcommand \bvec{\left( \begin{array}{c} }
\newcommand \evec{\end{array} \right)}

\newcommand \bea{\begin{eqnarray} }
\newcommand \eea{\end{eqnarray} }
\newcommand \nn {\nonumber}
\newcommand {\be} {\begin{equation}}
\newcommand {\ee} {\end{equation}}

\newcommand {\mbx} {\mbox{}}

\newcommand \ata {&\times&}

\begin{document}

\title{Modified Dihadron Fragmentation Functions in Hot and Nuclear Matter}

\author{A. Majumder}
\affiliation{Department of Physics, Duke University, Durham, NC 27708 }
\affiliation{Nuclear Science Division,
Lawrence Berkeley National Laboratory,
1 Cyclotron road, Berkeley, CA 94720}

\author{Enke Wang}
\affiliation{Institute of Particle Physics, Huazhong Normal University,
         Wuhan 430079, China}
\affiliation{Nuclear Science Division,
Lawrence Berkeley National Laboratory,
1 Cyclotron road, Berkeley, CA 94720}

\author{Xin-Nian Wang}
\affiliation{Nuclear Science Division,
Lawrence Berkeley National Laboratory,
1 Cyclotron road, Berkeley, CA 94720}

\date{ \today}

\begin{abstract}
Medium modification of dihadron fragmentation functions due to
gluon bremsstrahlung induced by multiple partonic scattering is studied
in both deep-inelastic scattering (DIS) off large nuclei and
high-energy heavy-ion collisions 
within the same framework of twist expansion.
The modified fragmentation
functions for dihadrons are found to follow closely that of single
hadrons leading to a weak nuclear suppression of their ratios as
measured by HERMES in DIS experiments. Meanwhile, a moderate
medium enhancement of the near-side correlation of two high transverse 
momentum hadrons with increasing 
centrality is found in heavy-ion collisions because of
the trigger bias and the increase in parton energy loss with centrality. 
Successful comparisons between theory and experiment for multi-hadron 
observables in both confining and deconfined environments offers comprehensive 
evidence for partonic energy loss as the mechanism of jet modification in dense matter.
\end{abstract}


\preprint{LBNL-56705}

\maketitle

Medium modification of the jet structure has emerged as a new
diagnostic tool for the study of partonic properties of the dense
matter~\cite{Gyulassy:2003mc}. The modification goes beyond a mere
suppression of inclusive spectra of leading hadrons and can be
extended to include the modification of many particle observables,
the simplest of which are two-hadron correlations within the
jet cone. Such two-hadron correlations have been measured
both in DIS~\cite{dinezza04} and high-energy heavy-ion
collisions~\cite{star2,phenix2}. While the two-hadron correlation
is found slightly suppressed in DIS off a nucleus versus
a nucleon target, it is moderately enhanced in central $Au+Au$
collisions relative to that in $p+p$. This is in sharp contrast
to the observed strong suppression of single inclusive
spectra~\cite{hermes1,highpt} in either DIS or central $A+A$ collisions and 
constitutes a necessary test of any energy loss formalism used to 
compute the modification of the single inclusive spectrum. 

Theoretically, multi-particle correlations from jet fragmentation
and their medium modification can be studied through $n$-hadron
fragmentation functions which can be defined as the overlapping
matrices of partonic field operators and $n$-hadron states. These
$n$-hadron fragmentation functions are non-perturbative and
involve long distance processes. However, they may be factorized
from the hard perturbative processes and their evolution with
momentum scale may be systematically studied in perturbative QCD
(pQCD) \cite{amxnw}, similarly as the usual single-inclusive
hadron fragmentation function (SFF) \cite{col89}. 
In this respect, 
they serve as an important test of the factorized pQCD formalism 
currently being applied in the computation of jet-like observables.

In this Letter, we report our first study of the medium modification
of dihadron fragmentation functions (DFF) in both cold nuclei and hot
deconfined partonic matter. The medium modification of DFFs
in DIS off nuclei will be derived
within the
framework of generalized factorization and twist expansion~\cite{Luo:1994np,guowang}.
The results are then extended to the case of
parton propagation
in heavy-ion collisions. With exactly the same
parameters determined from the modified SFFs
\cite{guowang,EW1}, the medium
modification of two-hadron correlations in both DIS off nuclei and
heavy-ion collisions are predicted and compared to the experimental data.

Similar to SFFs, the DFF of a quark into two hadrons of flavor 
$h_1,h_2$, with total forward momentum $p_h^+ \equiv p_1^+ + p_2^+$ 
(total momentum fraction $z \equiv z_1 + z_2$) and 
relative transverse momentum $q_{\perp}=p_{1 \perp}-p_{2 \perp}$ 
can be defined as the Fourier transform of the 
overlapping matrices of the quark fields $\psi_q (x),\bar{\psi}_q (0) $ 
and two-hadron inclusive final states $| p_1, p_2, S-2 \rc$ as,
\bea
D_q^{h_1,h_2}(z_1,z_2) &=& \frac{z^4}{4z_1z_2}
\int \frac{d^2q_\perp}{4(2\pi)^3}
\int \frac{d^4 p}{(2\pi)^4}
\nn \\ & & \hspace{-1.0in}\times
\kd \left( z - \frac{p_h^+}{p^+}  \right)
{\rm Tr} \Bigg[ \frac{\g^+}{2p_h^+}
\int d^4 x e^{i p \x x} \sum_{S - 2}
\nn \\ & &\hspace{-1.0in} \times
\lc 0 | \psi_q (x) | p_1, p_2, S-2 \rc
\lc p_1, p_2, S-2 | \bar{\psi}_q (0) | 0 \rc \Bigg],
\eea
and can be factorized from the hard processes~\cite{amxnw}.
They also satisfy the Dokshitzer-Gribov-Lipatov-Altarelli-Parisi (DGLAP)
evolution equations as 
derived in Ref.~\cite{amxnw}.
One unique feature of these equations
is the contribution from independent
fragmentation of two partons after the parton splitting.
These equations 
have been solved numerically~\cite{amxnw} and the $Q^2$ evolution
agrees very well with results
from JETSET~\cite{jetset} Monte Carlo simulations of
$e^+ + e^- \rightarrow h_1+h_2+ X$ processes. Although both
the single and dihadron fragmentation functions evolve rapidly
with $Q^2$, their ratio has a very weak $Q^2$ dependence.
In the absence of experimental data, JETSET Monte Carlo results will be used as
the initial condition for the vacuum DFF
in this study. For SFFs, 
the BKK parameterization~\cite{bin95} will be used; which also agrees
well with JETSET results.

Applying factorization to dihadron production in single jet events in
DIS off a nucleus, $e(L_1)+A(p)\rightarrow e(L_2)+h_1(p_1)+h_2(p_2) +X$,
one can obtain the dihadron semi-inclusive cross section,
\begin{equation}
E_{L_2}\frac{d\sigma^{h_1h_2}_{\rm DIS}}{d^3L_2 dz_1dz_2}
=\frac{\alpha^2}{2\pi s}\frac{1}{Q^4}L_{\mu\nu}\frac{dW^{\mu\nu}}{dz_1dz_2},
\end{equation}
in terms of the semi-inclusive tensor at leading twist,
\begin{eqnarray}
\frac{d W^{\mu \nu } }{d z_1 d z_2} &=&\sum_q \int d x  f^A_q(x,Q^2)
H^{\mu \nu}(x,p,q) \nonumber \\
&\times& D_q^{h_1,h_2}(z_1,z_2,Q^2).
\end{eqnarray}
In the above, 
$L_{\mu\nu}\!\!\!\!=\!\!\!(1/2){\rm Tr}(\not\!\!\!L_1\gamma_\mu\!\!\!\!\!\not\!\!\!L_2\gamma_\nu\!$),
the factor $H^{\mu \nu}$ represents the hard part of quark
scattering with a virtual photon which carries a four-momentum
$q=[-Q^2/2q^-,q^-,\vec{0}_\perp]$ and $f^A_q(x,Q^2)$ is the quark
distribution in the nucleus which has a total momentum
$A[p^+,0,\vec{0}_\perp]$. The hadron momentum fractions,
$z_1=p_1^-/q^-$ and $z_2=p_2^-/q^-$, are defined with respect to
the initial momentum $q^-$ of the fragmenting quark.

At next-to-leading twist~\cite{guowang,Luo:1994np}, 
the dihadron semi-inclusive tensor
receives contributions from multiple scattering of the struck
quark off soft gluons inside the nucleus with induced gluon
radiation. One can reorganize the total contribution
(leading and next-to-leading twist) into a product of effective
quark distribution in a nucleus, the hard part of photon-quark
scattering $H^{\mu \nu}$ and a modified DFF:
$\tilde{D}_q^{h_1,h_2} (z_1,z_2)$ (which includes, within it, 
the effect of subsequent scattering and gluon radiation). The calculation of the modified DFF  
at the next-to-leading twist in a nucleus proceeds~\cite{maj04f}
similarly as that for the modified 
SFFs~\cite{guowang} and yields,
\bea
\tilde{D}_q^{h_1,h_2} (z_1,z_2) &=& D_q^{h_1,h_2}(z_1,z_2) +
\int_0^{Q^2} \frac{dl_{\perp}^2}{l_{\perp}^2} \frac{\A_s}{2\pi} \nn \\
& &\hspace{-1in} \times\left[ \int_{z_1+z_2}^1 \frac{dy}{y^2}
\left\{ \D P_{q\ra q g} (y,x_B,x_L,l_\perp^2)
D_q^{h_1,h_2} \left(\frac{z_1}{y},\frac{z_2}{y} \right)\right.\right. \nn \\
& &\hspace{-0.8in}+\left. \D P_{q\ra g q} (y,x_B,x_L,l_\perp^2)
D_g^{h_1,h_2} \left(\frac{z_1}{y},\frac{z_2}{y} \right) \right\} \nn \\
& & \hspace{-0.8in}+\int_{z_1}^{1-z_2} \frac{dy}{y(1-y)}
\D \hat{P}_{q\ra q g} (y,x_B,x_L,l_\perp^2) \nn \\
& &\hspace{-0.8in}\times  \left. D_q^{h_1}
\left(\frac{z_1}{y})\right) D_g^{h_2}\left(\frac{z_2}{1-y} \right) + (h_1 \ra h_2) \right]\, .
\label{eq-dihdr-mod}
\eea
\nt
In the above, $x_B= -Q^2/2p^+q^-$, $x_L = l_\perp^2/2p^+q^-y(1-y)$,
$l_\perp$ is the transverse momentum of the radiated gluon, 
$\D P_{q\ra qg}$ and $\D P_{q\ra g q}$
are the modified splitting functions with momentum fraction $y$, 
whose forms are identical
to that in the modified 
SFF~\cite{guowang}.
The switch $(h_1 \ra h_2)$ is only meant for the last term, which
represents independent fragmentation of the quark and radiated gluon. 
The corresponding modified splitting
function,
\bea
\D \hat{P}_{q\ra gq} = \frac{1+y^2}{1-y}
\frac{C_A 2\pi \A_s T^A_{qg} (x_B,x_L)}{(l_\perp^2+\lc k_\perp^2\rc)
N_c f_q^A(x_B,Q^2)}, \label{splitting_func}
\eea
\nt
is similar to $\D P_{q\ra qg}$ but does not contain
contributions from virtual corrections.
In the above, $C_A=N_c=3$, and $\lc k_\perp^2\rc$ is the
average intrinsic parton transverse momentum inside the nucleus.

Note that both modified splitting functions depend on the
quark-gluon correlation function $T^A_{qg}$ in the nucleus that
also determines the modification of 
SFFs~\cite{guowang}. 
For a Gaussian nuclear distribution, it
can be estimated as,
\begin{equation}
T^A_{qg}(x_B,x_L) = \tilde{C}(Q^2) m_N R_A f_q^A(x_B) (1 - e^{-x_L^2/x_A^2}),
\label{eq-tqg}
\end{equation}
where, $x_A=1/m_NR_A$, $m_N$ is the nucleon mass and $R_A = 1.12
A^{1/3}$ is the nuclear radius. 
The small variation of the gluon correlation function is neglected;
its average value is absorbed into the overall constant
$\tilde{C}$. This is the only parameter in the modified DFF 
which might depend on the
kinematics of the  process but is identical to the parameter in
the modified SFF.
In the phenomenological study of the SFFs
in DIS off nuclei, $\tilde{C}=0.006$ GeV$^2$ was
determined within the kinematics of the HERMES experiment~\cite{hermes1}. 
The predicted dependence of nuclear modification
of the SFF
on the momentum
fraction $z$, initial quark energy $\nu=q^-$ and the nuclear size
$R_A$ agrees very well with the HERMES experimental data~\cite{EW1}. 

\begin{figure}[htbp]
\resizebox{2.4in}{2.4in}{\includegraphics[1.25in,0.75in][6.75in,6.25in]{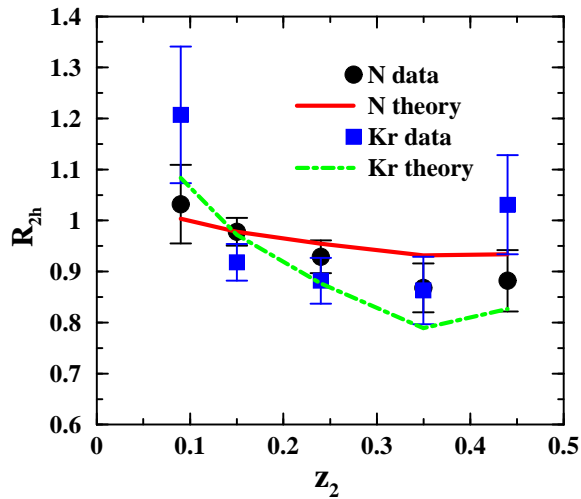}}
    \caption{ (Color online) Results on the medium modification of the associated hadron distribution 
[$R_{2h}(z_2)=N^A_{2h}(z_2)/N^1_{2h}(z_2) $, see Eq.~(\ref{eq-corr})] in nuclei versus its 
momentum fraction as compared with HERMES data \cite{dinezza04} in DIS with Nitrogen (N) and 
Krypton (Kr) targets.}
    \label{fig1}
\end{figure}

With no additional parameters in Eq.~(\ref{eq-dihdr-mod}),
one can predict the nuclear modification of 
DFFs within the same kinematics. Since 
the DFFs
are connected to 
SFFs via
sum rules~\cite{amxnw}, it is more illustrative to study the
modification of the distribution for the second rank hadrons normalized 
by the number of leading hadrons (\tie, hadrons with $z>0.5$),
\begin{eqnarray}
N_{2h}(z_2)\equiv \int_{0.5}^{1-z_2} dz_1D_q^{h_1,h_2}(z_1,z_2) \Bigg/ 
\int_{0.5}^1 dz_1 D_q^{h_1}(z_1),
\label{eq-corr}
\end{eqnarray}
where $z_1$ and $z_2<z_1$ are the momentum fractions of the
triggered (leading) and associated (secondary) hadrons, respectively.
Shown in Fig.~\ref{fig1} is the predicted ratio of the normalized associated
hadron distribution in DIS off a Nitrogen ($A=14$) and Krypton ($A=84$) target to
that off a proton ($A=1$) \tie, $R_{2h}(z_2)=N^A_{2h}(z_2)/N^1_{2h}(z_2)$, 
as compared to the HERMES experimental
data~\cite{dinezza04}.

The agreement between the prediction
and the data is remarkable given that no free parameters are used.
The suppression of 
$R_{2h}(z_2)$
at large $z_2$  with atomic number,
is quite small compared to the
suppression of the 
SFFs~\cite{hermes1,EW1}.
Since $N_{2h}(z_2)$ is the ratio of double and single hadron
fragmentation functions, the effect of induced gluon radiation or
quark energy loss is mainly borne by the single spectra of the leading
hadrons. At small values of $z_2$, the modified 
DFF rises above its vacuum counterpart more than the modified SFF. 
This is due to the new contribution where each of the
detected hadrons emanates from the independent fragmentation of
the quark and the radiated gluon. In the 
experiment, the measured $\nu$ and $Q^2$ vary with 
$z_1$ and $z_2$; this has been incorporated in the calculation


In high-energy heavy-ion (or $p+p$ and $p+A$) collisions, jets are
always produced in back-to-back pairs. Correlations of two
high-$p_T$ hadrons in azimuthal angle generally have two Gaussian
peaks \cite{star2,phenix2}. Relative to the triggered hadron,
away-side hadrons come from the fragmentation of the away-side jet
and are related to SFFs.
On the other hand, near-side hadrons come from the fragmentation of the
same jet as the triggered hadron and therefore
are related to DFFs.

The near-side correlation, background
subtracted and integrated over the azimuthal angle~\cite{fqwang,note},
can be related to the associated hadron distribution or
the ratio of the DFF to  the SFF,  both averaged over the initial jet energy weighted with the 
corresponding parton production cross sections. 
We report on the first such calculation in this Letter. 
Assuming a factorization of initial and final state effects (as done in the 
case of single inclusive observables~\cite{EW1,xnwang03}), 
the differential cross-section 
for the production of two high $p_T$ hadrons at midrapidity from 
the collision of 
two nuclei $A$ and $B$ at an impact parameter $b$ between $b_{min},b_{max}$ 
is given as, 

\bea
\frac{d \sigma^{AB}}{dy d {p_T^{\rm trig}} d {p_T^{\rm assoc}} } 
 &=&  \int_{\mbx_{b_{min}}}^{\mbx^{b_{max}}}\!\!\!\!\!\!\!\!\!d^2 b 
\int d^2 r t_A(\vec{r}+\vec{b}/2) t_B(\vec{r} - \vec{b}/2)  \nn \\
&\times& \!\!\!\! 2K \int \!\!\! d x_a d x_b  G^A_a(x_a,Q^2)  G^B_b(x_b,Q^2)  \nn \\
\ata \!\!\!\! \frac{  d \hat{\sg}_{ab \ra cd} }{ d \hat{t}}\tilde{D}_c^h(z_1,z_2,Q^2),  \label{AA_sigma}
\eea
\nt
where, $G^A_a(x_a,Q^2) [G^B_b(x_b,Q^2)]$ represents the nuclear parton 
distribution function for a parton $a (b)$ with momentum fractions $x_a(x_b)$ 
in a nucleus $A(B)$, $t_A$ ($t_B$) represents the 
nuclear thickness function and $d \hat{\sg}_{ab \ra cd} / d \hat{t}$ represents 
the hard parton cross section with Mandelstam variable $\hat{t}$. The final 
state momentum fractions $z_1,z_2$ represent the the momentum fractions of 
the two detected hadrons with respect to the initial jet energy. The 
factor $K \simeq 2$ accounts for higher order corrections 
(identical to that used in the case of the single inclusive spectra). 
The medium modified DFF 
may be expressed as in Eq.~\eqref{eq-dihdr-mod}, with the modified 
splitting functions generalized from Eq.~\eqref{splitting_func} as,

\bea
\D P_{q \ra i} &=& P_{q \ra i}(y)  2\pi \A_s C_A   
T^{M}(\vec{b},\vecr,x_a,x_b,y,l_\perp) \nn \\ 
\ata \left[\frac{\mbox{\Large}}{\mbox{\Large}} \right. l_\perp^2 N_c 
 t_A(\vec{r}+\vec{b}/2) t_B(\vec{r} - \vec{b}/2) \nn \\ 
 \ata \left.  G^A_a(x_a) G^B_b(x_b) 
\frac{d \hat{\sg}}{d \hat{t}} \right] ^{-1}  +  v.c. \label{TM}
\eea

\nt
As in the case of DIS, $l_\perp$ represents the transverse momentum of the 
radiated gluon and $T^{M}$ replaces the factor $T_{qg}^A$ (in Eq.~\eqref{splitting_func})and represents 
the parton-gluon correlation function in a quark-gluon plasma. The primary difference between the denominators of Eqs.~\eqref{splitting_func} and~\eqref{TM}, lies in the weighting 
of a particular jet like parton by the initial production process
(see Ref.~\cite{Majumder:2006we} for the evaluation of $T^M$ in single inclusive observables).

\begin{figure}[htbp]
  \resizebox{2.4in}{2.4in}{\includegraphics[0.5in,0.4in][4.2in,4.1in]{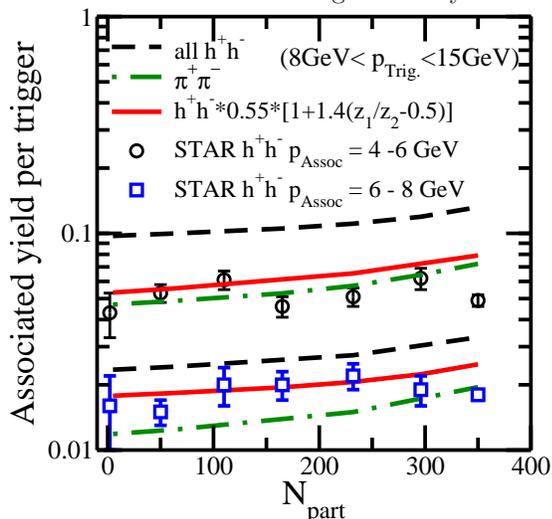}}
\caption{ (Color online)
Yields of different flavors of hadrons with 
6 GeV $<p_T^{\rm assoc.} < p_T^{\rm trig.}$ and 4 GeV $ < p_T^{\rm assoc.} < 6$GeV, associated with a trigger 
hadron with $p_T^{\rm trig.} > 8$ GeV versus the centrality  of  $Au$+$Au$ collisions at $\sqrt{s}=200$ GeV 
as compared to experimental data \cite{Adams:2006yt}, see text for details. }
 \label{AA_2}
\end{figure}

At present, there exists no measurement of high momentum 
hadrons associated with a high $p_T^{\rm trig.}(>8$GeV) hadron in the case of $p$+$p$ collisions. 
Therefore, we will compare the  associated yields in $d$+$Au$ 
and $Au$+$Au$ 
collisions as a function of centrality as well as the associated $p_T$.
These yields are 
very sensitive to the number of flavors detected. In Fig.~\ref{AA_2}, the 
yield of charged hadrons with associate transverse momentum in two ranges  
(6GeV$<\!p_T^{\rm assoc.}\!\!<\! p_T^{\rm trig.}$ and 4GeV$ < \!p_T^{\rm assoc.}\!\!< 6$GeV),  associated with a 
hadron with $p_T^{\rm trig} > 8$GeV, as measured in Ref.~\cite{Adams:2006yt}, 
are presented along with three different calculations using 
Eqs.~(\ref{AA_sigma}) and (\ref{TM}). 
The dashed lines 
correspond to the case of all charged hadrons where as the dot-dashed 
lines corresponds to the case of charged pions only. In both cases, the 
vacuum DFFs are estimated from JETSET Monte Carlo 
simulations (see Ref.~\cite{amxnw}). Unlike in the case of the 
experimental measurements, no decay corrections have been introduced 
in these calculations:
the number of pairs measured, includes not only those produced 
directly from the fragmentation of a jet but also decay products of 
particles from fragmentation. As a result, such fragmentation functions 
tend to be somewhat larger at lower momentum fractions than fragmentation 
functions with such corrections. The integrated (over transverse momentum)
associated yield being dominated by lower momentum fractions shows a larger 
effect as lower momentum ranges in $p_T^{\rm assoc.}$ are chosen. 

A decay corrected DFF [$\bar{D}(z_1,z_2)$] may be constructed, phenomenologically,  by comparing with the 
differential spectrum of associated particles in $d$+$Au$ collisions 
(as a function of $z_T=p_T^{\rm trig.}/p_T^{\rm assoc.}$, in Ref.~\cite{Adams:2006yt}): 
\bea
\bar{D}(z_1,z_2) \!\!&=&\!\! 0.55 \times D(z_1,z_2) [1 + 1.4(z_1/z_2 - 1 )] .
\eea
\nt
Using the above DFF, we obtain a very good agreement with the 
data (solid lines). 
All estimates show a mild rise with centrality which originates solely from the 
increased trigger bias as the centrality of the collision is 
increased. 
We have focused 
on large $p_T^{\rm trig.}$ and $p_T^{\rm assoc.}$, as large momenta ensure the validity
of the independent fragmentation picture. Otherwise, other non-perturbative
and higher twist effects such as recombination~\cite{recom} and the influence of
radial and longitudinal flow can become important
and may lead to further modification of the associated yield~\cite{rec2}.  
As may be noted from Fig.~\ref{AA_2}, lowering the $p_T^{assoc.}$
results in a systematic departure between prediction and data with increasing centrality.

In summary, we have studied the modification of the DFFs 
in both confining and hot deconfined matter due to 
multiple parton scattering and induced gluon radiation. The modification, 
follows closely that of SFFs so that the associated hadron distributions or the ratios
of DFFs to SFFs are only slightly
suppressed in DIS off nuclei but enhanced in central heavy-ion
collisions due to trigger bias. With no extra parameters, our
calculations agree very well with experimental data. Such calculations, combined with previous successful 
comparisons in the single inclusive sector~\cite{EW1} in both cold nuclear and hot deconfined matter constitutes 
a very important test of the partonic jet modification formalism and testifies 
to the applicability of perturbatively calculable jet observables as 
probes of the dense matter created in heavy-ion collisions.

We thank F. Wang for discussions.
This work was supported by the U.S. Department of Energy under 
Contract No. DE-AC03-76SF00098, under grant DE-FG02-05ER41367, NSFC under project Nos. 10475031 
and 10135030 and NSERC of Canada.


\begin{thebibliography}{99}
\bibitem{Gyulassy:2003mc}
M.~Gyulassy {\it et al.}, 
arXiv:nucl-th/0302077;
X.~N.~Wang,
arXiv:nucl-th/0405017.

\bibitem{dinezza04}
A.~Airapetian {\it et al.},
  Phys.\ Rev.\ Lett.\  {\bf 96}, 162301 (2006).

\bibitem{star2} C.~Adler {\it et al.}, 
Phys.\ Rev.\ Lett.\  {\bf 90}, 082302 (2003).

\bibitem{phenix2}
S.~S.~Adler {\it et al.}, 
 Phys.\ Rev.\  C {\bf 71}, 051902 (2005).



\bibitem{hermes1}
A.~Airapetian {\it et al.}, 
Eur.\ Phys.\ J.\ C {\bf 20}, 479 (2001);
Phys.\ Lett.\ B {\bf 577}, 37 (2003).

\bibitem{highpt}
K.~Adcox {\it et al.}, 
Phys.\ Rev.\ Lett.\  {\bf 88}, 022301 (2002);
C.~Adler {\it et al.}, 
Phys.\ Rev.\ Lett.\  {\bf 89}, 202301 (2002).

\bibitem{amxnw}
A.~Majumder and X.~N.~Wang,
Phys.\ Rev.\ D {\bf 70}, 014007 (2004);
Phys.\ Rev.\  D {\bf 72}, 034007 (2005);
U.~P.~Sukhatme, and K.~E.~Lassila,
Phys. \ Rev. \ D. {\bf 22}, 1184 (1980).

\bibitem{col89}
J.~C.~Collins, D.~E.~Soper and G.~Sterman,
Adv.\ Ser.\ Direct.\ High Energy Phys.\  {\bf 5}, 1 (1988).

\bibitem{Luo:1994np}
  M.~Luo, J.~W.~Qiu and G.~Sterman,
  Phys.\ Rev.\ D {\bf 50}, 1951 (1994).

\bibitem{guowang} X.~F.~Guo and X.~N.~Wang,
Phys.\ Rev.\ Lett.\  {\bf 85}, 3591 (2000);
X.~N.~Wang and X.~F.~Guo,
Nucl.\ Phys.\ A {\bf 696}, 788 (2001);
J.~Osborne and X.~N.~Wang,
Nucl.\ Phys.\ A {\bf 710}, 281 (2002);
B.~W.~Zhang and X.~N.~Wang,
Nucl.\ Phys.\ A {\bf 720}, 429 (2003).

\bibitem{EW1}
E.~Wang and X.~N.~Wang,
Phys.\ Rev.\ Lett.\  {\bf 89}, 162301 (2002);
{\it ibid.},
{\bf 87}, 142301 (2001).

\bibitem{jetset}
B.~Andersson {\it et al.} , 
Phys.\ Rept.\  {\bf 97}, 31 (1983);
T.~Sjostrand,
arXiv:hep-ph/9508391.


\bibitem{bin95}
J.~Binnewies, B.~A.~Kniehl and G.~Kramer,
Phys.\ Rev.\ D {\bf 52}, 4947 (1995).



\bibitem{maj04f}
A.~Majumder and X.~N.~Wang, {\it to be published}.



\bibitem{fqwang}
F.~Wang, 
J.\ Phys.\ G {\bf 30}, S1299 (2004)

\bibitem{note}
Jet structure in $p+p$ collisions is known to have correlated
background due to multiplicity bias and coherence in 
hadronization. This develops a non-trivial centrality and 
azimuthal angle dependence in $p+A$
and $A+A$ collisions. Ref.~\cite{star2} uses the two-hadron correlation
in $p+p$ collisions, including background, as the baseline,  
while Ref.~\cite{fqwang} subtracts out the background both 
in $p+p$ and $A+A$ collisions. Such different methods of 
background subtraction affect the extracted dihadron correlations 
introducing systematic differences.


\bibitem{xnwang03}
X.~N.~Wang,
Phys.\ Lett.\ B {\bf 595}, 165 (2004).


\bibitem{Majumder:2006we}
  A.~Majumder,
Phys.\ Rev.\  C {\bf 75}, 021901 (2007).

\bibitem{Adams:2006yt}
  J.~Adams {\it et al.},
 Phys.\ Rev.\ Lett.\  {\bf 97}, 162301 (2006).

\bibitem{recom}
R.~C.~Hwa and C.~B.~Yang,
Phys.\ Rev.\ C {\bf 67}, 034902 (2003);
R.~J.~Fries {\it et al.}, 
Phys.\ Rev.\ Lett.\  {\bf 90}, 202303 (2003);
V.~Greco, C.~M.~Ko and P.~Levai,
Phys.\ Rev.\ Lett.\  {\bf 90}, 202302 (2003).

\bibitem{rec2}
  R.~C.~Hwa and C.~B.~Yang,
  Phys.\ Rev.\ C {\bf 70}, 054902 (2004);
  R.~J.~Fries, S.~A.~Bass and B.~Muller,
  Phys.\ Rev.\ Lett.\  {\bf 94}, 122301 (2005).




\end{thebibliography}
\end{document}